\documentclass[a4paper]{article}

\usepackage[utf8]{inputenc}
\usepackage[T1]{fontenc}
\usepackage[english]{babel}
\usepackage[protrusion=true,expansion=true]{microtype}
\usepackage{graphicx}

\usepackage{subfig}
\usepackage{xcolor}
\usepackage{amsmath}
\usepackage{cancel}

\usepackage{cite}
\usepackage{listings}
\usepackage{authblk}

\usepackage{hyperref}

\newcommand*\diff{\mathrm{d}}

\title{A method to challenge symmetries in data with self-supervised learning}
\author{Rupert~Tombs}
\author{Christopher~G.~Lester}
\affil{
Cavendish~Laboratory,
University~of~Cambridge,
CB3~0HE,
United~Kingdom
}

\date{July 2022}

\begin{document}

\maketitle

\section*{Abstract}
\noindent
Symmetries are key properties of physical models and of experimental designs,
but any proposed symmetry may or may not be realized in nature.
In this paper, we introduce a practical and general method to test such
suspected symmetries in data, with minimal external input.
Self-supervision, which derives learning objectives from data
without external labelling, is used to train models to predict `which is real?'
between real data and symmetrically transformed alternatives.
If these models make successful predictions in independent tests, then they
challenge the targeted symmetries.
Crucially, our method handles filtered data, which often arise from
inefficiencies or deliberate selections, and which could give the illusion of
asymmetry if mistreated.
We use examples to demonstrate how the method works and how the models'
predictions can be interpreted.

Code and data are available at \url{https://zenodo.org/record/6861702}.

\section{Introduction}
If observations of nature refute a symmetry, then all models that imply that
symmetry are wrong.
Particularly in fundamental physics, important theoretical models and
experimental designs do possess symmetries, but we are not always certain
whether nature actually reflects those symmetries.
Testing suspected symmetries in data is therefore of significant
scientific value, since refuting a symmetry also indirectly challenges all
physical hypotheses which imply it.

In this paper, we introduce a method which can perform such tests on a general class of
symmetries for any form of computerized data, and without external labels or descriptions of those data.
Implementing our method requires only data,
knowledge of any filtering on those data,
a way to transform those data according to the candidate symmetry,
and an appropriate learning algorithm to model the asymmetries.

This method, which we name `which is real?', begins by building pairs of data, each containing
one real datum and one `fake' copy that has been transformed according to the
proposed symmetry.
If we then build a machine that reliably predicts which entry in each pair is real,
and which is fake, then the symmetry must be broken.
We build and test such machines with modern learning algorithms according
to standard machine learning practice: train with one portion of the data,
and generate reliable results by testing in another independent portion.

Real data are messy.
Not only do data take complicated shapes in various digital representations,
they are usually also subject to lossy effects in collection and processing.
Such inefficiencies, however, are often well understood, and that understanding
is productively leveraged to select subsets of the data which have cleaner
connections to their idealized modelling.
To test symmetries in practice, it is therefore crucial that our method works
seamlessly in the presence of such lossy effects or selections,
which we describe collectively as filtering the data.

Note that asymmetry in data does not necessarily imply asymmetry in
the fundamental physics of nature.
Data arise from natural and artificial effects, and a proposed symmetry could
be broken in any step in the chain of data preparation,
such as hardware construction or software analysis.
This is not necessarily a defect of our method;
evidence for asymmetry indicates that either nature is asymmetric
or that our experiments need recalibrating, and both lessons
can be valuable.

\paragraph{Context}
Our `which is real?' method uses an example of self-supervised learning,
in which one trains models to perform tasks derived from the data themselves
without external labels ---
with external labels, it would be supervised learning.
In common practice, self-supervision helps to teach models about structures in
the data, such that they later become more useful for other tasks~\cite{
dosovitskiy2014iscriminative,
multitaskself2017,
oord2019representation,
devlin2019bert,
chen2020simple,
grill2020bootstrap,
pmlr-v139-radford21a
}.
Our emphasis is not on the training, however, but on the models' performance
with the self-supervised `which is real?' task itself.

Active research pursues ways to build symmetries into machine learning
algorithms~\cite{
pmlr-v48-cohenc16,
maron2019invariant,
thomas2018tensor,
zaheer2018deep,
finzi2021practical%
}, with scientific applications including high energy physics
\cite{
komiske2019energy,
shmakov2021spanet,
favoni2020lattice,
kanwar2020equivariant,
dillon2021symmetries,
hepmllivingreview,
lester2021stressed%
}
and astronomy
\cite{
scaifePorter2021,
dieleman2015rotation
}.
Building symmetries into model architectures usefully encodes domain-specific
prior information, and so improves performance by leaving less to learn from data.
In contrast, this paper seeks to challenge those symmetric priors in contexts
where they may or may not hold in nature.

If a symmetry of interest is considered to be normal or expected behaviour,
then our method has an application in the automated detection of collective
anomalies~\cite{chandola2009anomaly}.
The study of anomaly detection has a substantial history in and outside of particle
physics, including a modern emphasis on the use of machine learning
tools~\cite{
guansong2021deep,
desai2022symmetry,
hepmllivingreview
}.

Our method is a probabilistic recasting of ideas developed in a sister
paper~\cite{lester2021stressed}, which uses neural networks to approach similar
tasks in the absence of filtering effects.
We informally discuss some mathematical properties of symmetries that
are formalized in another sibling paper~\cite{lester2021chiral}.
We have also demonstrated the `which is real?' method, which is introduced in this paper,
as a tool to test for anomalous parity violation in simulated particle physics
data~\cite{lester2022hunting}.

\paragraph{Layout}
Before detailing the `which is real?' method, we introduce
what we mean by a symmetry and the notation we use to talk about it
in Section~\ref{sec:symmetries}.
We use Section~\ref{sec:is-it-real} to describe how a standard classification
setup could alternatively be used to test symmetries,
but also how it would struggle to handle filtered data.
`Which is real?' avoids this struggle; it is described in
Section~\ref{sec:which-is-real} and demonstrated with examples:
a cylindrical particle detector in Section~\ref{sec:ex-ring},
and a landscape with less trivial symmetry in Section~\ref{sec:ex-map}.
We briefly discuss extensions beyond the core method in
Section~\ref{sec:extensions}, and Section~\ref{sec:summary} is a terse summary.

\section{Symmetries}
\label{sec:symmetries}
\begin{figure}[t]
\subfloat[]{
\includegraphics[height=0.24\textwidth]{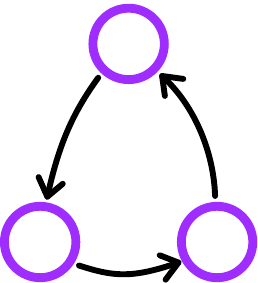}
\label{fig:symmetries-cycle}
}
\hfill
\subfloat[]{
\includegraphics[height=0.24\textwidth]{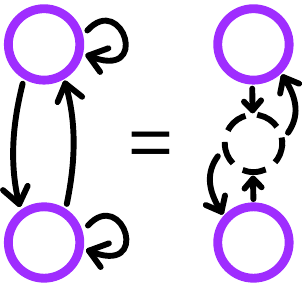}
\label{fig:symmetries-orbit}
}
\hfill
\subfloat[]{
\includegraphics[height=0.24\textwidth]{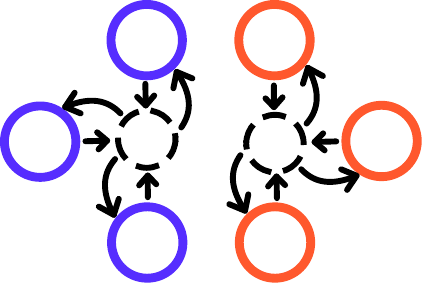}
\label{fig:symmetries-two-orbits}
}
\caption{%
Some symmetries of discrete distributions.
(a)~A deterministic transition which cycles between three elements
of a distribution;
if each element has equal probability, then this transition
is a symmetry since it only rotates the probability masses between the atoms.
(b)~A discrete distribution over two elements with an orbital symmetry
which randomly transitions from each element to itself or to its neighbour.
The dashed ring is an alternative illustration of the same transition,
showing how each state transitions into the orbit, then back out to specific states
within that orbit.
(c)~A discrete distribution with six elements partitioned into two similar orbits.
}
\label{fig:symmetries}
\end{figure}

Without a precise definition of what we mean by a symmetry, the remainder
of this paper would be quite unclear.
This section provides general, measure-theoretic definitions for symmetries
as we think of them here, along with our conventional notation.

Wishing to capture any operation which has no observable effect on the
distribution of data, we define a symmetry of a distribution to be any
transformation which leaves that distribution unchanged.
That is, although the transformation should change individual data,
transforming all elements of a distribution together must preserve
all statistical properties of the ensemble.%
\footnote{%
Under this definition, every non-trivial distribution has very many symmetries.
For example, a discrete distribution with $n$ equiprobable elements has $n!$
permutation symmetries, and every continuous distribution can be transformed
to uniform density in appropriate coordinates, in which every rearrangement
is a symmetry.
Shuffled cards and stirred drinks are examples.
This is why we begin from candidate symmetries proposed for scientific interest,
and do not attempt to discover interesting symmetries from first principles.
}

In notation, we consider data states $x$
with probability distributions $p(x)$.
A candidate symmetry $S$ defines a transformation $\tau_S(x'\mid x)$
which takes real data $x$ to new, fake data $x'$.
For generality, $\tau_S(x'\mid x)$ is a probability distribution:
it may randomly sample alternatives $x'$,
but can still specify non-random transitions by leveraging Dirac delta distributions.
Transforming all elements of $p$ through $\tau_S$ induces a
new distribution $p_S$, and we define that $S$ is a symmetry of $p$
if and only if $p_S = p$.

To illustrate this definition of a symmetry, Figure~\ref{fig:symmetries} displays
examples in small discrete distributions; Figure~\ref{fig:symmetries-cycle}
illustrates a deterministic symmetry that cycles probability mass between states,
and Figure~\ref{fig:symmetries-orbit} illustrates a random transition
between two elements.
Parity symmetry, $S_\mathrm{P}$, is a physically important example that uses the operation
$x \rightarrow x' = -x$, and could define the deterministic transition distribution
\begin{equation}
\label{eqn:tau-parity}
\diff \tau_{S_\mathrm{P}}\!(x'\mid x) = \delta(x' + x) \,\diff x'
.
\end{equation}
Continuous rotational symmetry on a circle, $S_\circ$ with $x \in [0, 2\pi)$,
could be encoded by a random transition,
\begin{equation}
\label{eqn:tau-circle}
\diff \tau_{S_\circ}\!(x'\mid x)
= \frac{1}{2\pi} \boldsymbol{1}_{0 \leq x' < 2\pi} \,\diff x'
,
\end{equation}
which happens to be independent of $x$.
(We include the $\diff x'$ terms as reminders of the coordinates in which these
density functions live and the Jacobian factors that are necessary to change coordinates.)

Although we do not attempt to formalize exactly how a conceptual symmetry should
correspond to a transition function, we do note that there is an implication structure
among transformations.
For example, if a distribution is symmetric under all rotations, then it is also
symmetric under rotations by $\pi$.
So if a test indicates asymmetry under the transformation of rotating by $\pi$,
then it also indicates violation of the more general rotational symmetry.

Analytically, the induced distribution for given $p$ and $\tau_S$ is calculable as
\begin{equation}
\label{eqn:ps_integral}
p_S(x') = \int \tau_S(x'\mid x) \,\diff p(x)
;
\end{equation}
the probability mass at each $x'$ is a mixture of all elements in the
original distribution $p$, weighted by their transitions through $\tau_S$.
The methods of this paper, however, use empirical distributions only,
so do not require the direct computation of this integral.
In matrix notation, this integral is $p_S = \tau_S p$, where
$\tau_S$ is a left-stochastic transition matrix; our definition of a
symmetry coincides with $p$ being the stationary state of a Markov chain
with transition matrix $\tau_S$.

\paragraph{Measure theory}
Although we prefer to avoid excessive formality, the above definitions may be
formalizable in the language of measure theory:
$p$ is a probability measure, and $p_S$ is its pushforward through
a measurable function $\tau_S$.
Our definition of $S$ being a symmetry of the distribution $p$ then coincides
with $p$ being an invariant measure of $\tau_S$.

Defining a measure-theoretic symmetry in this way naturally covers both continuous
and discrete cases, and avoids any need for densities with respect to given
coordinates, which alternative formulations could
consider~\cite{desai2022symmetry}.
Measure theory also provides `Radon-Nikodym' derivatives between measures, such as
\begin{equation}
\frac{\diff p}{\diff p_{S\!\!}}(x),
\end{equation}
which is the relative density of $p$ with respect to $p_S$,
and does not require intermediate coordinates against which to define that
density~\cite{billingsley2008probability}.
Symmetry ($p_S = p$) means that $(\diff p/\diff p_S)(x) = 1$,
and so our symmetry tests are essentially searches for examples where this
derivative differs from unity.

\paragraph{Orbits}
Our definition of a symmetry is very general, and the `which is real?' method
works most efficiently for important special cases that we name the orbital symmetries.
Not only do orbital symmetries permit a valuable simplification, they matter in practice because
they include all of our intended applications.
We define and explore orbital symmetries here, and will see their benefit to the
`which is real?' method in Section~\ref{sec:which-is-real}.

Orbital symmetries are symmetries which partition the data space into subsets,
called orbits, within which all accessible states share a common transition distribution;
to illustrate this concept, Figures~\ref{fig:symmetries-orbit}
and~\ref{fig:symmetries-two-orbits} display orbital symmetries in two
simple discrete systems.

For example, rotational symmetry about a $(\theta, z)$ cylinder has an orbit for
each $z$, at which $\tau_S$ could be a distribution over new points $(\theta', z)$
with uniform density in $\theta'$, independent of the original $\theta$.
The circular example of Equation~\ref{eqn:tau-circle} is an orbital symmetry
with one orbit covering all data, and parity symmetry can be tested as an orbital symmetry
if its transition distribution is adapted from Equation~\ref{eqn:tau-parity} to a sum of
deltas at both $+x$ and $-x$.

Group symmetries are particularly important in physics, and therefore orbital symmetries
are also physically important because every group symmetry is an orbital symmetry.

To show this, we first define each transition function of a group symmetry to be a
distribution over the set accessible by operating with elements of a group on the
given datum.
More precisely, suppose that each $x$ comprises a group element $x_g$ with
some baggage $x_b$ that identifies a group $G_b$.
That is, $x = (x_g, x_b)$ with $x_g \in G_b$.
Then the operations in the group map $x$ to the unordered set
$\mathrm{Orbit}(x) = \{(g \cdot x_g, x_b): g \in G_b\}$
which equals $\{(g, x_b): g \in G_b\}$
(up to a permutation by Cayley's theorem, but we are not interested in order).
The set is therefore independent of $x_g$ within $G_b$, satisfying our definition
of an orbital symmetry.

Finally, we note that the transition distribution of an orbital symmetry does not
depend precisely on the original datum $x$, but on the orbit to which it belongs;
that is, an orbital $\tau_S(x' \mid x)$ depends only on $x'$ and $\mathrm{Orbit}(x)$,
and therefore only on $x'$ since $\mathrm{Orbit}(x) = \mathrm{Orbit}(x')$.
This reduced dependency simplifies the `which is real?' method, and the general
applicability to group symmetries gives us confidence in its practical use.

\section{Standard classification}
\label{sec:is-it-real}
For standard binary classification, a dataset contains data of two classes
that are sampled from two distributions, and the task of a classification model
is to predict the class labels of given data.
One can test symmetries in such a classification setup by constructing two
classes, one of `real' data that are observed from $p$,
and another of `fake' data that we sample from the transformed distribution $p_S$;
the key to our methods is that if an algorithm successfully classifies these
real and fake data, then the two distributions must differ, and the symmetry must be broken.

How should success be quantified?
Perhaps $95\%$ classification accuracy sounds like a great success,
but if the dataset contains $95\%$ real data, then it is actually a failure since naive guessing would do equally well.
More precisely, then, our classifier is only successful if it does \emph{better}
than an alternative that assumes parity symmetry.
Our classifiers model asymmetry between the distributions,
and we compare them against that alternative
symmetric hypothesis through likelihood ratios on their label predictions.

We can only train and test example classifiers, but predictions from the symmetric hypothesis are in fact uniquely determined: since symmetry requires $p = p_S$, the symmetric hypothesis
learns nothing from the data, and can only assign all label probabilities equally
by their proportions in the distribution at large.
For example, if we choose to sample real and fake data in equal proportions, then the symmetric
model always assigns probability $1/2$ to each label.
To predict labels better than this, our asymmetric classifiers must
use information from the data themselves.

The remainder of this section discusses how such classifiers can be trained,
how we report their likelihood ratios,
and how this standard approach unfortunately struggles to handle data filtering.

\paragraph{Odds}
Our machine learning tools are general function approximators that construct
functions with values on the real line, but we use them to assign label probabilities,
and probabilities are awkwardly constrained between $0$ and $1$.
It is therefore practical (and entirely standard) to first approximate $\log$ odds ratios with these functions,
then convert those to probabilities. We develop that conversion here.

A $\log$ odds function $\phi(x)$ relates to the data distributions as
\begin{equation}
\label{eqn:real-fake-log-odds}
\phi(x)
=
\log \frac
{p(\textrm{real} \mid x)}
{p(\textrm{fake} \mid x)}
=
\log \frac{\diff p}{\diff p_{S\!\!}}(x)
+
\log
\frac
{p(\textrm{real})}
{p(\textrm{fake})}
,
\end{equation}
which is comfortably unbounded on the real line, so suitable for learning.
Note that to derive this relationship, we have used the alternative notation of
$p(x) = p(x \mid \textrm{real})$ and $p_S(x) = p(x \mid \textrm{fake})$.
Similarly, the prior label probabilities $p(\textrm{real})$ and $p(\textrm{fake})$
are the relative the rates with which one samples from $p(x)$ and $p_S(x)$ respectively.

We choose to use equal parts real and fake: $p(\textrm{real}) = p(\textrm{fake}) = 1/2$.
In general, however, these rates are a free choice that could be tuned
to each application;
they control both the distribution of the data,
which affects how well algorithms will learn,
and the distribution of labels,
which affects the precision in testing.
For example, $p(\textrm{fake}) > 1/2$ could help by providing more training
data in regions with more fakes, but the extreme choice that $p(\textrm{fake}) = 1$
would provide no information in testing.

Given an assigned $\log$ odds function $\phi(x)$ and
equal label proportions, some probability algebra recovers that
\begin{equation}
\label{eqn:logistic}
p(\textrm{real}\mid x) = 1 / (1 + e^{-\phi(x)})
,
\end{equation}
which is well known as the logistic function, and that maps our learned
$\log$ odds to a likelihood function over labels.

Learning can follow the gradients of a loss function defined in the standard way
as the negative mean $\log$ likelihood on a batch of $n$ data, which we write as
\begin{equation}
\label{eqn:loss-standard}
\mathcal{L}(\{\ell_i\}, \{x_i\}, \phi)
=
\frac{1}{n}\sum_{i=1}^{n}
\left\{
\begin{matrix}
\log(1 + e^{-\phi(x_i)}) &\textrm{if}~\ell_i=\textrm{real},~\textrm{and} \\
\log(1 + e^{+\phi(x_i)}) &\textrm{if}~\ell_i=\textrm{fake}\hphantom{,~\textrm{and}} \\
\end{matrix}
\right.
,
\end{equation}
where $\ell_i \in \{\mathrm{real}, \mathrm{fake}\}$ are the labels on data.

\paragraph{Model comparison}
Data contribute to model comparison through ratios of likelihoods,
which in this case are ratios of the probabilities assigned to the labels on data.
To quantify these ratios in a manner that can converge towards a fixed value
as the number of data $n$ increases, we define a `quality' function $Q$
as the mean $\log$ likelihood ratio:
\begin{equation}
\label{eqn:quality}
Q(\{\ell_i\}, \{x_i\}, p)
= \frac{1}{n}\sum_{i=1}^{n} \log p(\ell_i \mid x_i) - \log \frac{1}{2}
\end{equation}
where the learned asymmetric model assigns $p(\ell_i \mid x_i)$,
which compare against the symmetrical assignment of $1/2$.
Evidently, this relates to the loss function of
Equation~\ref{eqn:loss-standard} by $Q = \log 2 - \mathcal{L}$.

This quality function $Q$ may be recognized as a difference of binary
cross-entropies~\cite{MurphyKevinP.2012Mlap},
and is invertible to the likelihood ratio for known $n$.
(So $n$ should always be reported!)
With perfect predictions, the model attains $\mathcal{L}=0$ and the
maximum $Q$ of $\log2 \approx 0.69$,
which is the expected loss of the symmetric hypothesis,
and equivalently the entropy of the binomial likelihood
that we controlled (and maximized) by choosing equal label probabilities.
Finally, $Q$ is not bounded from below, since a bad model can approach
assigning zero probability to an observed label, and that mistake is
severely punished in the loss function.

Since the model does not change during testing, the summands in $Q$ are independent,
and standard estimation methods can assign reasonable uncertainties to its
limiting value.
To inform uncertainty judgements, we therefore report $Q$ with the standard deviation
of its summands per $\sqrt{n}$.

If $Q$ (more precisely, the corresponding $\log$ likelihood ratio)
is large and positive, then the model is predicting more
accurately than symmetry would allow, indicating that the symmetry is
violated.
If, however, the data obey the symmetry, then one should expect $Q$ to be negative, since
learning algorithms are unlikely to find the ideal $p(\ell \mid x) = 1/2$ solution.
Similarly, small $Q$ does not confirm the symmetry, but indicates that
it is not refuted by this attempt with these data and this learning algorithm.

\paragraph{Filtering and weighty problems}
Without filtering, it is straightforward to prepare datasets
with real and fake data sampled from $p$ and $p_S$ respectively:
just transform random subsets of the data under $\tau_S$, and label them
appropriately.

Practical experiments, however, usually do not record all possible data,
due to holes, inefficiencies, or selections which filter the data before analysis.
We describe filtering effects with a function $L(x)$,
which assigns a probability that data generated at $x$ are accepted.
For example, an unreliable sensor might have $L(x) = 0.1$,
and hard removal under a selection requirement would have $L(x) = 0$.

We assume in this work that $L(x)$ is exactly known.
If it is not, then a fixed approximation should be assigned at training time,
and although such an approximation may not be optimal, it may be practical.
In testing, however, an inaccurately approximated filter could falsely indicate asymmetry
if mistreated, so filter uncertainties should be respected in the testing phase.
We do not fully develop the handling of filter uncertainties in this paper,
but review them briefly in Section~\ref{sec:extensions}.

With filtering, data are not observed directly from $p(x)$ but from the filtered
distribution $f(x) \propto L(x)p(x)$.
Since we are interested in symmetries of data distributions, not of filters,
fakes should be sampled as if pure data were collected from $p(x)$,
resampled from $\tau_S(x' \mid x)$, and then filtered by $L(x')$.
Applying these relations, the filtered distribution of fakes is
\begin{equation}
\label{eqn:filtered_fakes}
f_S(x')
\propto \int
\frac{L(x')}{L(x)}
\tau_S(x'\mid x) \,\diff f(x)
.
\end{equation}
Since we can only approximate $\diff f(x)$ with an ensemble of real filtered data,
the simplest strategy to sample this distribution would be to transform each
entry through $\tau_S(x'\mid x)$ and weight it by something proportional
to this ratio of filter functions.

Although that ratio cannot cause division-by-zero errors,
since no data can be observed with $L(x) = 0$,
we do have a problem when data survive tight filtering,
leading to unboundedly large weights when $L(x') \gg L(x)$.
In extreme cases, large weights can dominate the estimated fake distribution,
giving an unappealing imbalance and poor statistical precision.
This is the primary problem with weights on data,
which we use the `which is real?' method to avoid.

A secondary problem of unbounded weights is that they cause
a little dependency between data.
Independence is valuable because rigorous testing requires independent data on which to test,
and because learning from independent mini-batches is efficient:
computationally, to reduce data transfer rates,
and algorithmically, given the empirical successes of stochastic learning.
Weights must be normalized to fix the label proportions
$p(\textrm{real})$ and $p(\textrm{fake})$,
and the problem arises because their normalization constants can only be
estimated from finite data.
Suppose, for example, that we estimate normalization with a running average
that updates as each mini-batch is consumed in training.
When a new, highly weighed observation is included,
then it should suddenly change the weights on all other data, in all batches,
in a clearly non-independent fashion.
Although we expect that normalization from moderately sized batches
would normally be practical, this risk of domination by extreme outliers
leaves a little awkwardness.

Many cases will have well-behaved weight distributions, and this standard
classification strategy can be used effectively in practice.
For these cases and others, the `which is real?' method tests symmetries
similarly, but with the benefits that it acts without weights
and with full independence between data.

\section{Which is real?}
\label{sec:which-is-real}
Rather than comparing two classes in a mixed distribution,
the `which is real?' method tests symmetries through a self-supervised task of
discriminating each data entry from a transformed `fake' clone of itself,
and avoids weights by not attempting to undo any filtering,
but by applying the filter a second time when sampling the fake.

For each datum $x$, `which is real?' samples one fake $x'$ from the transition
distribution $\tau_S(x' \mid x)$ to form an $x'\textrm{--}x$ pair,
and from that pair, the method derives two labels corresponding to its two
possible orderings:
fake--real ($x$ is real and was transformed to $x'$; $x' \leftarrow x$),
and
real--fake ($x'$ is real and was transformed to $x$; $x \leftarrow x'$).

Unlike standard classification, the choice of fake--real and real--fake
proportions does not change the data distribution, so we have no reason
to construct them in unequal proportions and choose
$p(\textrm{fake--real})$ = $p(\textrm{real--fake}) = 1/2$ always.
In this way, we extract from each datum a binary bit which discriminates itself
from its fake clone, and use that bit as the gem of information to be predicted
by a classifier.

Since the classification problem is now to predict which label,
fake--real or real--fake, is correct,
a naive and approach could learn a label probability on the joint
$x'\textrm{--}x$ space in the standard way.
However, that would mean working with an input space of pairs
(comprising real and fake data together)
that are twice as large as the original data,
and that can easily incur costs in both learning performance and interpretability.
For orbital symmetries, however, we show that the problem shrinks to learning
a smaller function $\zeta(x)$ on the original data space, and which assigns probabilities through differences $\zeta(x) - \zeta(x')$.

\paragraph{Classification}
To clarify the following analysis, we use $p(x' \leftarrow x)$ as shorthand for
$p(x'\textrm{--}x \mid \textrm{fake--real})$
which corresponds to the observation of a real $x$
that was subsequently transformed into $x'$ to produce the pair $x'\textrm{--}x$.
Given this sequential construction, the joint probability of a fake--real ordered
data pair factors into the steps of observation and transformation:
\begin{equation}
p(x' \leftarrow x) = \tau_S(x' \mid x) p(x)
.
\end{equation}
From this, we construct a $\log$ odds over real--fake and fake--real ordering labels,
as we did for standard classification.
Unlike standard classification, however, that $\log$ odds
now splits into an exchange-odd difference of two similar terms, since
\begin{align}
\phi(x', x)
&= \log \frac{
\diff p(x' \leftarrow x)
}{
\diff p(x \leftarrow x')
}
\label{eqn:transition-ratio-p}
\\
&=
\log \frac{\diff p}{\diff \tau_{S\!\!}}(x \mid x')
- \log \frac{\diff p}{\diff \tau_{S\!\!}}(x' \mid x)
\\
&= \xi(x, x') - \xi(x', x)
.
\vphantom{\frac{1}{1}}
\label{eqn:transition-ratio-xi}
\end{align}
To be explicit, $(\diff p/\diff \tau_S)(x \mid x')$ is the
relative density of $p(x)$ with respect to $\tau_S(x \mid x')$,
and $\xi(x, x')$ is the learnable function which should approximate
its logarithm up to an additive constant.
We have also dropped the prior label probabilities,
which vanish because we have assumed them to be equal.

Although $\xi(x', x)$ depends on the (big) joint space of $x'$ and $x$,
that shrinks if we consider only orbital symmetries.
For an orbital symmetry, any accessible $x\textrm{--}x'$ pair lives within one orbit,
and $\xi(x', x)$ does not depend directly on $x$, but only on the orbit to
which both $x'$ and $x$ belong.
Since we can deduce the orbit from either
$x$ or $x'$, we can replace $\xi(x', x)$ for an orbital symmetry with a new
learnable function $\zeta(x')$, and use the simpler assignment
\begin{equation}
\label{eqn:phi_zeta}
\phi(x', x) = \zeta(x) - \zeta(x')
.
\end{equation}
This construction from a difference ensures that $\phi(x, x')$ is correctly
antisymmetric under exchange of its arguments, meaning that there is no
difference between
$x\textrm{--}x'$ labelled real--fake
and
$x'\textrm{--}x$ labelled fake--real;
both cases give identical results, so there is no reason to shuffle the
orderings.
We therefore choose to always order all pairs fake--real, without cost.

Again, the logistic function of Equation~\ref{eqn:logistic} relates the
$\log$ odds to the label probability, and the loss function to minimize
in training for the `which is real?' task is again a negative mean $\log$ likelihood
\begin{equation}
\label{eqn:wir-loss}
\mathcal{L}(\{(x'_i, x_i)\}, \zeta)
=
\frac{1}{n}\sum_{i=1}^{n} \log (1  + e^{-(\zeta(x_i) - \zeta(x_i'))})
,
\end{equation}
which is recognized as the binary cross-entropy for classification
in the special case that its label is constant, and that its $\log$ odds
is exchange-odd between the $x$ and $x'$ inputs.
Following its definition in Equation~\ref{eqn:quality}, the corresponding
quality is $Q = \log 2 - \mathcal{L}$, as before.
Note that pairs with $x = x'$ are constant in this loss function, so they
can optionally be excluded for efficiency.

\paragraph{Filtering}
To handle filtering, we now generate fakes not by undoing
the initial filter, but by applying it a second time after the transition,
leading to the filtered transition distribution
\begin{equation}
f_S(x' \mid x) \propto L(x') \tau_S(x' \mid x)
.
\end{equation}
Sampling from such an arbitrary distribution can be challenging,
but practical strategies exist:
most simply, one can follow its construction by sampling from $\tau_S(x' \mid x)$
and accepting each sample with probability $L(x')$;
this is known as `rejection sampling'.
Rejection sampling can be inefficient when $L(x')$ rejects most samples,
but efficiency can be improved by adapting the initial sampling towards
$f_S$, and again rejecting a proportion of events to correct for
remaining differences.
Taking this further, ideal adaptation is achieved though
`inverse transform sampling' in which one inverts a cumulative distribution
function to map from the unit uniform distribution onto $f_S$ with no rejections.

Viewed as Bayesian computation, this is an ordinary case of posterior sampling
from a prior $\tau(x' \mid x)$ constrained by a likelihood $L(x')$,
so various Markov Chain Monte Carlo algorithms are available
for the most challenging cases~\cite{
sivia2006data,
roberts2004general,
neal2003slice,
neal2011mcmc
}.

With this choice to sample fakes in proportion to $L(x') \tau_S(x' \mid x)$,
the same filter function applies to both real and fake data. Therefore,
\begin{equation}
f(x' \leftarrow x) = L(x') \tau_S(x' \mid x) L(x) p(x)
,
\end{equation}
and all four instances of the filter cancel in the filtered $\log$ odds:
\begin{equation}
\label{eqn:phifiltered}
\phi(x, x')
= \log
\xcancel{\frac{L(x') L(x)}{L(x) L(x')}}
\frac{
\diff p(x' \leftarrow x)
}{
\diff p(x \leftarrow x')
}
,
\end{equation}
leading back to Equation~\ref{eqn:transition-ratio-p} as if there were
no filtering.
Learning now proceeds as before, with the loss function of
Equation~\ref{eqn:wir-loss}, in ignorance of any filtering except
in how it changes the available distribution of data.
This means that any successful learning algorithm should continue to
approximate the unfiltered result wherever it receives enough data.

\paragraph{Wrap-up}
Symmetry testing with `which is real?' works as for the more general case of
standard classification:
train a model with one portion of the data, and test it on another by
evaluating $Q$ according to Equation~\ref{eqn:quality}.
Now, however, the model attempts to discriminate each data entry from a faked
version of itself that is sampled according to the product of the
transition distribution with a filter function.

For orbital symmetries, learning produces a function $\zeta(x)$,
from which differences give the fake--real
$\log$ odds, which translate to probabilities through
the logistic function, whose logarithmic mean gives the differentiable
loss function of Equation~\ref{eqn:wir-loss} for use in training.
We do not specify which learning algorithms should use that loss function
because the method itself is indifferent; in any application, it is the task
of the user to apply a practical algorithm that is appropriate to their context,
be it a linear model, neural network, decision tree or something completely different.

The following examples exercise the `which is real?' method in some toy environments
that include much of the complexity of real data.
Standard machine learning tools perform well on these examples with minimal tuning,
although this is no surprise given the examples' simple nature.

\section{Example 1: Cylindrical detector}
\label{sec:ex-ring}
\begin{figure}[t]
\centering
\includegraphics[width=0.99\textwidth]{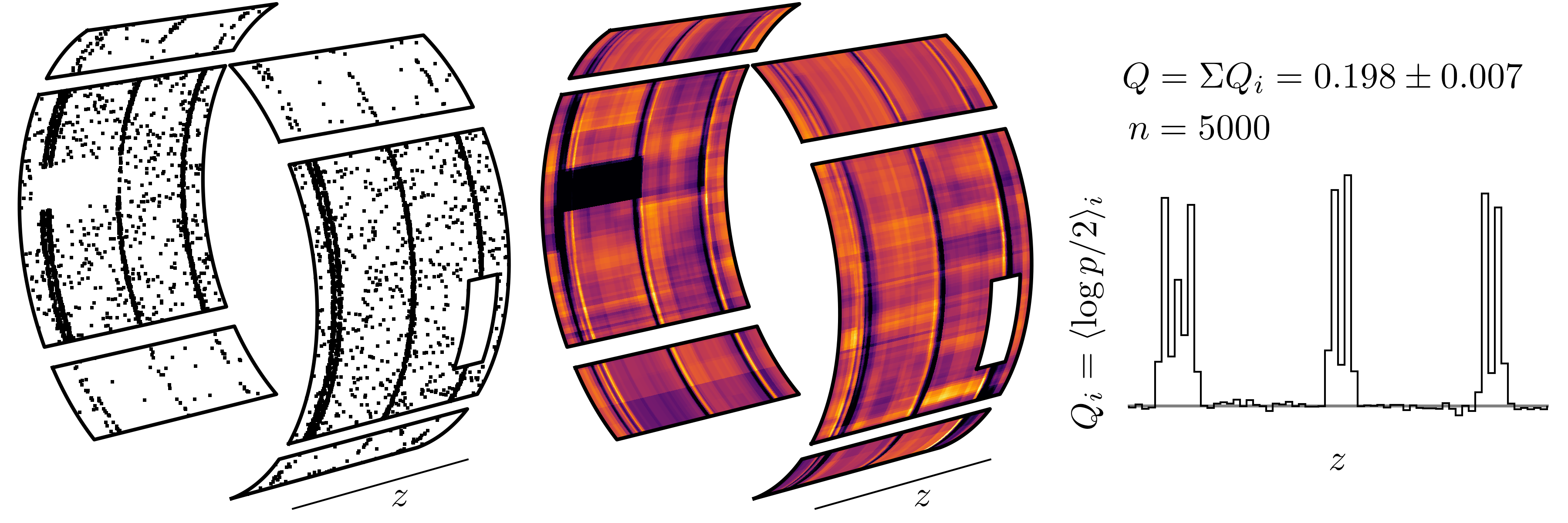}
\caption{%
(left)~Training data on a gappy cylindrical detector. Smaller
panels detect only $10\%$ of data in their regions. Gaps detect no data.
(middle)~A learned function predicts that data in bright yellow areas are more
likely to be real than versions in the dark purple regions;
it finds a missing patch and an offset which varies with angle.
The colour map ranges from $\zeta(x) = -2$ (black) to $+2$ (yellow).
(right)~Positive $Q$ with large $n$ is evidence against rotational symmetry;
the histogram shows contributions to this evidence from rotational orbits
along the length.
}
\label{fig:ring-abstract}
\end{figure}

\begin{figure}[t]
\centering
\subfloat[]{
\includegraphics[width=0.99\textwidth]{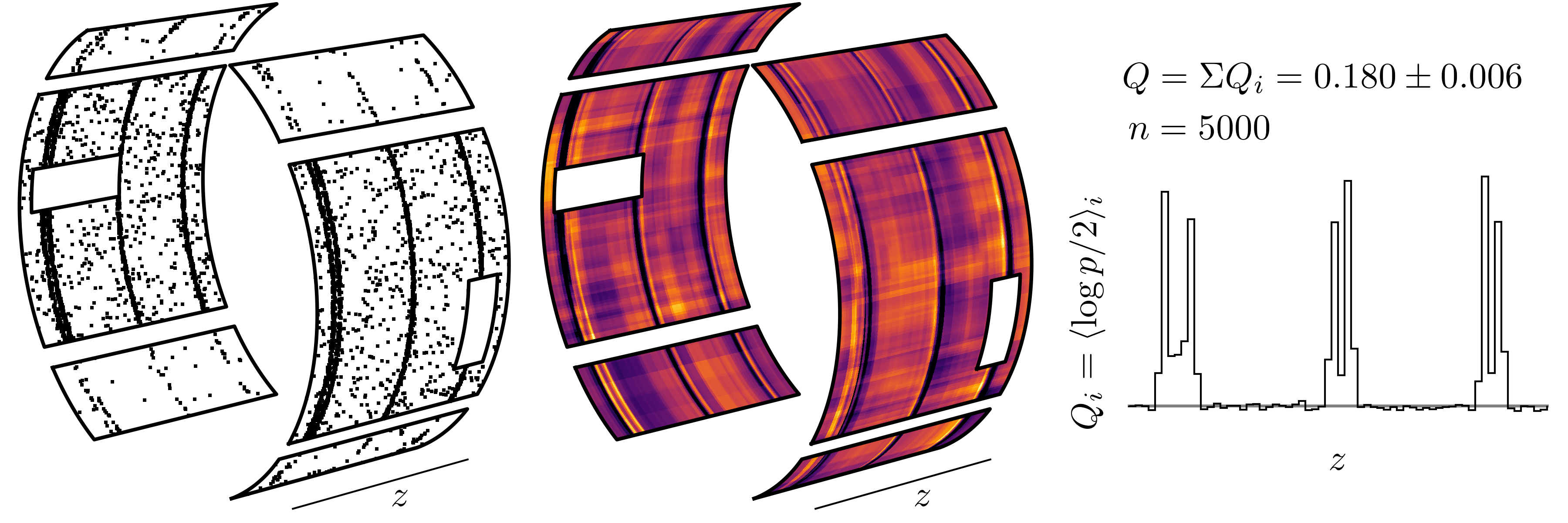}
\label{fig:ring-fix-hole}
}
\\
\subfloat[]{
\includegraphics[width=0.99\textwidth]{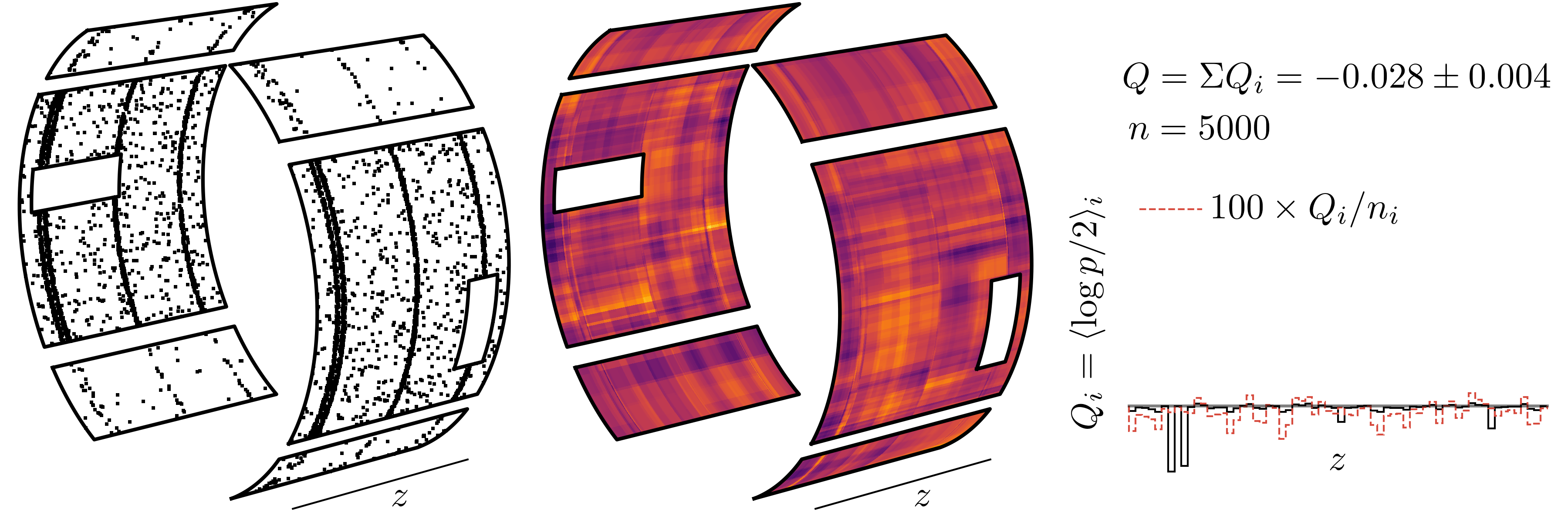}
\label{fig:ring-no-wave}
}
\caption{%
Modifications to the gappy cylindrical particle detector from
Figure~\ref{fig:ring-abstract}.
(a)~The hole in the back panel is included in the filter function,
weakening the observed asymmetry.
(b)~The sinusoidal offset is removed, and fresh data are sampled.
Since the asymmetric model assigns non-$1/2$ probabilities to data from
a symmetric distribution, it does worse than the symmetric model.
Here, negatively spiking $Q_i$ bins indicate where data are denser;
the dashed orange line shows the mean in each bin times $100$,
demonstrating that the model does not perform worse on average,
but that these spikes result from the accumulation of many data.
More details are given in the text of Section~\ref{sec:ex-ring}.
}
\label{fig:ring-example}
\end{figure}

To illustrate a filtered cylindrical symmetry, as might arise in particle
detector experiments, we design a broken ring that is illustrated in
Figure~\ref{fig:ring-abstract}.
This detector is imagined to record point data
on its surface, which could be estimates of the points where energetic particles
interacted with sensitive materials.

The visible holes in this detector imply a clear form of filtering in which
all data are lost.
We additionally design that the smaller panels miss $90\%$ of possible data,
perhaps because of their construction from thinner material.
The filter function is therefore $L(x) = 1$ for $x$ in a large panel,
$L(x) = 0.1$ for $x$ in a small panel, and $L(x) = 0$ otherwise in the gaps.

The candidate symmetry is of rotation about the axis of the cylinder.
We describe data by their longitudinal position $z$ and angle
$\theta$ from the vertical axis, so $x = (\theta, z)$,
and assign a transformation distribution $\tau_{S_\circ}\!(x' \mid x)$
that is uniform for $\theta$ between $0$ and $2\pi$,
as in Equation~\ref{eqn:tau-circle}, around orbits of fixed $z$.

For this example, we have assigned a non-trivial data distribution $p(x)$, but do not
describe it in detail here since the purpose of this method is to test for
symmetries in data without needing to know the details of their distribution.
We specify distributions and filters in discrete coordinates,
and sample them with discrete inverse transform sampling to choose pixels,
followed by direct uniform sampling within those pixels.

With these models and methods, we sample two sets of $5000$ fake--real
data pairs for training and testing respectively,
from which Figure~\ref{fig:ring-abstract}(left) shows a scatter
plot of the training data.

From these training data, we learn $\zeta(x)$ functions with LightGBM~\cite{lightgbm}%
\footnote{%
We use LightGBM~3.3.2 with default parameters through its scikit-learn
interface~\cite{scikit-learn, sklearn-api}, except for
\lstinline{subsample=0.5}, \lstinline{subsample_freq=1}, and the custom
`which is real?' objective of Equation~\ref{eqn:wir-loss}.
Without subsampling, the exactly repeated $z$ coordinates appear to hinder
training.
}
since it is a robust and efficient learning algorithm.
Like other boosted decision tree models~\cite{xgboost}, its output
is a weighted sum of step functions acting on trees of inequalities in the input coordinates,
for which the algorithm learns weights and locations from the first two
derivatives of the loss function of Equation~\ref{eqn:wir-loss},
with respect to $\zeta(x)$ and $\zeta(x')$.

A $\zeta(x)$ function learned by LightGBM is displayed in the middle plot of
Figure~\ref{fig:ring-abstract}.
In this figure, a large black hole has appeared where training data are lacking;
there, the model has learned that any data are probably fake, having been rotated in
from other angles.
Circular rings of high data density are visible in the scatter plot, and these rings match
with dark purple and bright yellow fringes that swap sides between the front and back panels.
This suggests an apparent misalignment between those dense rings and the axis of the cylinder,
as if data in these rings on the front panel are shifted to the left in $z$,
and those on the back panel are shifted to the right.

In the small panels, where data are sparse due to filtering, $\zeta(x)$ is
smoother but shows similar structures to the main panels;
despite the filtering in these small panels, the algorithm has still learned
towards the same underlying structure, but does so with less precision
where it has fewer data.

Test results are displayed in the rightmost plot of
Figure~\ref{fig:ring-abstract}, in which the histogram shows contributions
to the $Q$ sum (of Equation~\ref{eqn:quality}) from bins along the $z$ axis,
each of which accumulates contributions from orbits around $\theta$ in its range.
This figure shows positive contributions in bins lined up with the edges of the
dense rings in data; in these locations,
there are many data, and the fakes identify themselves due
to angular variations in the rings that are breaking rotational symmetry.
The accumulated total of
$Q = 0.198 \pm 0.007$ strongly indicates violation of rotational symmetry,
since it corresponds to a huge $\log$ likelihood ratio of $nQ = +990$ in favour
of the learned asymmetric model.

After considering the results of this first analysis, we re-examine our
imaginary detector and decide that the patch with no data in the back panel is a
gap in the hardware, much like the known hole on the front panel,
and so it was an error to not include it in the modelled filter function.

For a second phase, we therefore include that hole as a patch of zeros
in the filter function and repeat the experiment.
Results from this repeat are displayed in Figure~\ref{fig:ring-fix-hole},
and they correctly indicate a reduced asymmetry due to our correction of this
one source of experimental error.
Elsewhere, the $\zeta(x)$ function has not substantially changed;
this is a positive sign for the method, since it indicates that the filter
is being handled sensibly.
The remaining positive contributions to $Q$, however, indicate the presence
of other sources of asymmetry.

Thirdly, we imagine adjusting some knobs to recalibrate the alignment of our
detector, perhaps with respect to a beampipe, and repeat the experiment once more.
Results from this third test are displayed in Figure~\ref{fig:ring-no-wave},
and now show no sign of asymmetry;
the learned model receives a negative $Q$, saying that the symmetric hypothesis
of $p(\textrm{fake--real}\mid x) = 1/2$ is slightly more accurate.

The histogram to the right of Figure~\ref{fig:ring-no-wave} shows negative
contributions to $Q$ in spikes at the dense rings.
This is not because the model has learned poorly there, but
because those rings are dense, so contain more data to accumulate in the sum.
To demonstrate this, we overlay an orange-dashed histogram showing the average
contribution per data entry, and that average is in fact closer to
zero in these dense rings than elsewhere.

For this third test, the data were in fact sampled from a rotationally symmetric distribution
with the dense rings aligned exactly around the cylinder,
so the learned asymmetric model was doomed to fail.

\section{Example 2: Height map}
\label{sec:ex-map}
\begin{figure}[t]
\centering
\includegraphics[width=0.6\textwidth]{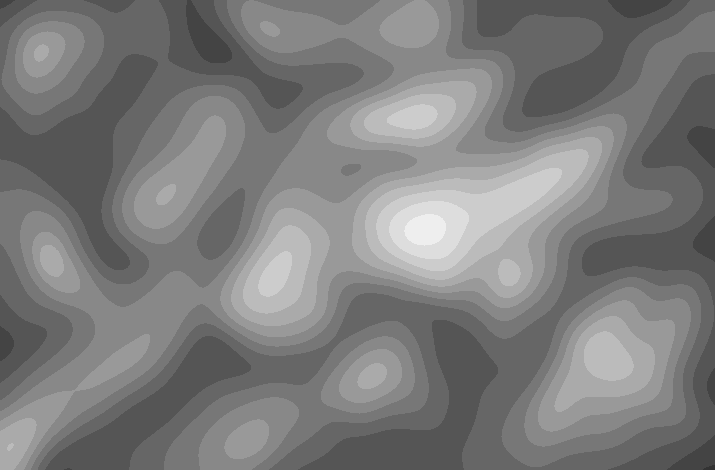}
\caption{%
Symmetrical contours: the height map symmetry of Section~\ref{sec:ex-map}
asserts that the data are translationally invariant within each contiguous
blob of constant brightness in this image.
That is, the terrain comprises discrete jumps and flat planes, as if it
were made of stacked foam boards or voxels in a computer game.
Data displayed in Figure~\ref{fig:map-zero} are sampled with density
that increases with the brightness of this landscape.
}
\label{fig:map-height}
\end{figure}

\begin{figure}[t]
\centering
\subfloat[]{
\includegraphics[width=0.98\textwidth]{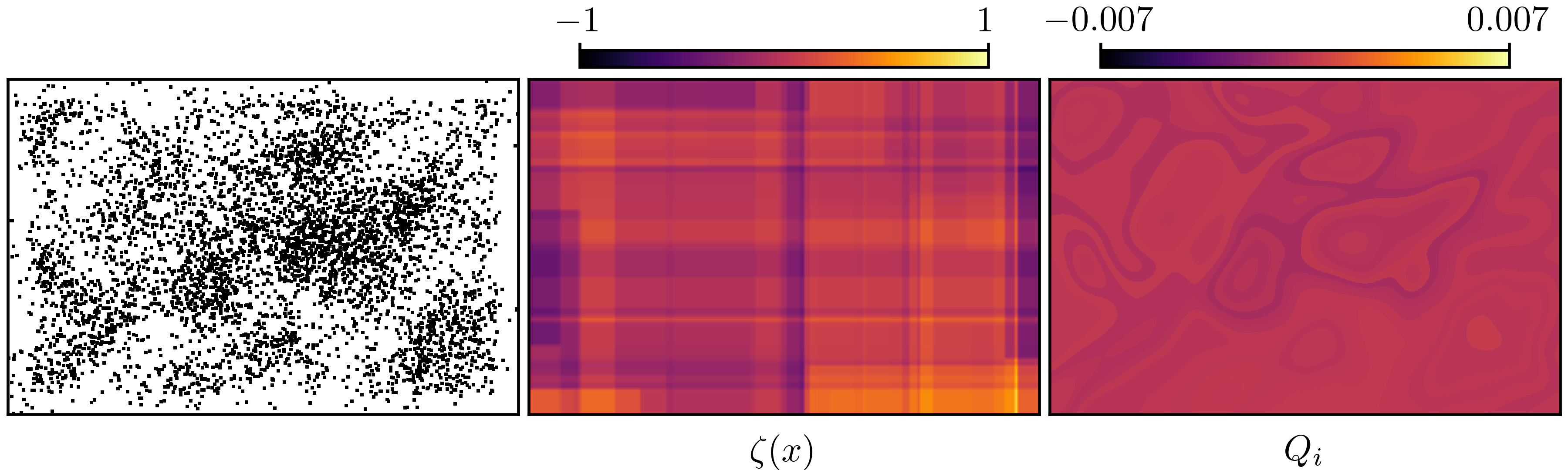}
\label{fig:map-zero}
}
\\
\subfloat[]{
\includegraphics[width=0.98\textwidth]{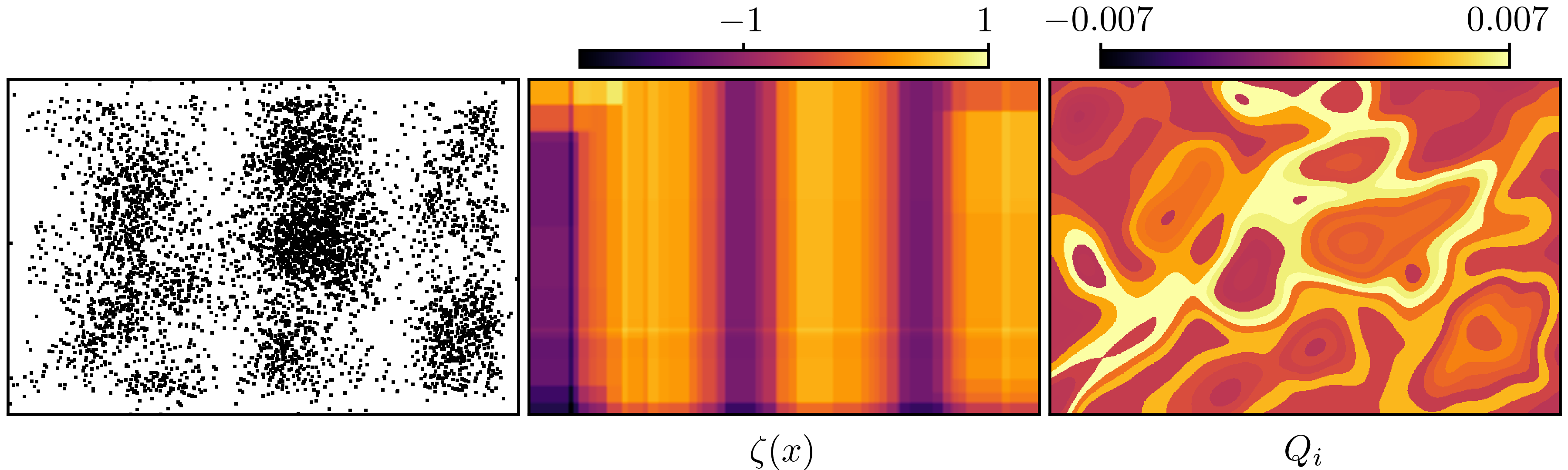}
\label{fig:map-nine}
}
\caption{%
Results of applying the `which is real?' method to data with the height-map
symmetry from Figure~\ref{fig:map-height}.
Data in the visible bordering band pass the filter with $20\%$
probability.
(a)~The sampling density of data increases with height-map brightness and
obeys the height-map symmetry.
For these symmetric data, $Q = -0.002 \pm 0.001$, so the model has not
learned exact symmetry and does worse than the symmetric model.
(b)~The data distribution violates the height-map symmetry
since its density is scaled by sinusoidally varying waves.
The model has learned these waves in its $\zeta(x)$ function, and
the largest contributions to $Q$ come from contours which contain many data and
have orbits which span large variations of those waves.
For these asymmetric data, $Q = 0.066 \pm 0.004$.
In both cases, $n=5000$.
More details are given in the text of Section~\ref{sec:ex-map}.
}
\label{fig:map-example}
\end{figure}

For cylindrical symmetry, and other common examples of orbital symmetries,
symmetric transformations reduce to simply re-sampling one or more natural
coordinates of $x$.
But this need not be the case; symmetries and their orbits may be non-trivial.
To demonstrate this, we propose a translational symmetry within contours of
constant height on a topographical map that is Figure~\ref{fig:map-height}.
In this figure, each contiguous area of constant brightness defines an orbit,
within which we test translational symmetry with the `which is real?' method
by assigning $\tau_S(x' \mid x)$ to be uniform in the plane within each orbit.

For this example, we simulate two experiments with data sampled from different
distributions.
The first experiment has data density which is uniform in the plane within each orbit,
and which increases with brightness, so that it does obey the symmetry.
The second experiment violates the symmetry by taking the data density from this
first experiment and scaling it by a sinusoidal function to produces waves along
the horizontal axis.
Plausibly justified by sparse observational coverage towards the edges of the map,
we implement a filter that removes $80\%$ of data in a narrow bordering region of
each experiment.

As in the example of Section~\ref{sec:ex-ring}, we use LightGBM models%
\footnote{
We use LightGBM~3.3.2 with default parameters except for
\lstinline{max_depth=2} (to reduce complexity) and the custom objective
of Equation~\ref{eqn:wir-loss}.
},
and sample two batches of $5000$ data for training and testing
by inverse transform sampling into pixels, followed by uniform sampling within
each chosen pixel.

Data and results from these experiments are displayed in Figure~\ref{fig:map-example}.
The first, symmetric experiment is shown in Figure~\ref{fig:map-zero}, and
achieves a negative $Q$, correctly indicating no evidence for asymmetry.
As anticipated, the model has imperfectly learned from noise in the finite data.
In the second experiment, those sinusoidal waves with which we broke the symmetry
are visible in Figure~\ref{fig:map-nine}, both in the scatter plot of data and
in the learned $\zeta(x)$ function.
Here also, the method succeeds by indicating asymmetry with a large positive $Q$.

On the right of this figure, we see that the largest contributions to $Q$ are
attributable to orbits containing both many data and large asymmetry.
In particular, that asymmetry is driven by oscillations of the waves we introduced,
so orbits with too little horizontal extent see less asymmetry than those
that span more than half a wavelength or so.

\section{Extensions}
\label{sec:extensions}
Although we described the `which is real?' task with only one fake per data
entry, multiple fakes can also be used by including them additively in the loss function.
While we previously averaged the loss function over $n$ data, if we now repeat each
entry $k$ times with newly sampled fakes, then we should average their $n\times k$ terms into
\begin{equation}
\label{eqn:wir-loss-k-fakes}
\mathcal{L}(\{(x'_{i,j}, x_i)\}, \zeta)
=
\frac{1}{nk}\sum_{i=1}^{n}\sum_{j=1}^k \log (1  + e^{-(\zeta(x_i) - \zeta(x_{i,j}'))})
.
\end{equation}
Using multiple fakes can reduce testing variance, particularly for discrete
symmetries with few equivalent states, but might help or hinder training
since stochastic optimization algorithms can benefit from additional noise.
Adding multiple fakes does add computational costs, but implementations can
be efficient since $\zeta(x)$ need only be evaluated once for the real entry
and once for each fake.

Uncertainty in the filter function is also important since uncertainty is likely to occur
whenever $L(x)$ is not deliberately introduced, and inaccuracies will impact test results
if our methods learn about errors in the filter function.
Filter uncertainty means we have two different filter functions in play:
one approximate filter used to simulate fakes, and one
imperfectly known filter that acted on the data before their observation.
With uncertainty, training can continue with the fixed approximate filter, since
although filter inaccuracies may lead to suboptimal learned models,
we do not require that models are optimal.
Any model that predicts successfully in rigorous testing comprises evidence of asymmetry,
so it is the rigorous testing that must respect the filter uncertainty,
and this means that the $\log$ odds should be modified to
allow for variations in the ratios between these two filters.

Introducing different filter functions for data and fakes leads to new, additive
terms in the $\log$ odds of Equation~\ref{eqn:phifiltered}, and those terms
should be allowed to vary according to some described uncertainty on the filter.
In testing, the accumulated likelihood over all predicted labels then depends on
those variations, for both the symmetric hypothesis and our learned asymmetric models.
That is, we have likelihood functions that vary under uncertain prior predictions;
while interesting, this is a standard and context-dependent data-analysis
problem~\cite{sivia2006data}, which we do not develop further here.

\section{Summary}
\label{sec:summary}
We have introduced a practical and general method for challenging symmetries in data,
which uses machine learning algorithms to train models to perform the self-supervised
task of answering `which is real?' between real data and `fake' transformed clones.
The method seamlessly handles data that are subject to known filtering effects,
and avoids weighting data with that filter by applying it twice,
once to real data and once to fakes.
Where the trained models answer successfully on independent testing data, they
reveal where they have found asymmetry and indicate its scale.
Success on the non-trivial examples of this paper demonstrates that the method is
ready to challenge theoretically or experimentally proposed symmetries in real
applications.

\section*{Acknowledgements}
We thank Peterhouse for hosting and feeding this endeavour.
RT thanks Daniel Noel for many useful questions and discussions,
the Cambridge \textsc{Atlas} machine learning discussion group,
Lars Henkelmann,
Homerton College,
Zelna Weich,
the Science and Technology Facilities Council,
and the Cambridgeshire weather for blowing away distractions.
We owe many thanks to the anonymous reviewer, whose comments have greatly
improved this paper.

\bibliography{bib}{}
\bibliographystyle{unsrturl}

\end{document}